\documentclass[10pt,twocolumn,letterpaper]{article}

\usepackage{wacv}
\usepackage{times}
\usepackage{epsfig}
\usepackage{graphicx}
\usepackage{amsmath}
\usepackage{amssymb}
\usepackage{amsmath}

\DeclareMathOperator*{\argmin}{arg\,min}

\wacvfinalcopy 

\newcommand{\nj}[1]{\textcolor{black}{#1}}
\newcommand{\sam}[1]{\textcolor{black}{#1}}
\newcommand{\wacvadded}[1]{\textcolor{black}{#1}}

\def\wacvPaperID{253} 

\ifwacvfinal\fi
\begin{document}


\title{Image Enhancement by Recurrently-trained Super-resolution Network}

\author{SAEM Park\\
Seoul National University, Seoul, Korea\\
LG Eletronics, Korea\\
{\tt\small parksaem2@snu.ac.kr}\\
\and
Nojun Kwak\\
Seoul National University, Seoul, Korea\\
{\tt\small nojunk@snu.ac.kr}
}

\maketitle
\ifwacvfinal\thispagestyle{empty}\fi

\begin{abstract}
    We introduce a new \nj{learning strategy for image enhancement by recurrently training the same \nj{simple} super-resolution (SR)} network multiple times. After \nj{initially training an SR network by using pairs of a corrupted low resolution (LR) image and an original image}, \nj{the proposed} method makes \nj{use of the trained SR network to generate} new high resolution (HR) images \nj{with a doubled resolution from the} original uncorrupted images. \nj{Then, the new HR images are downscaled to the original resolution, which work as target images for the SR network in the next stage}. \nj{The newly generated} HR images \nj{by the repeatedly trained SR network show} better \nj{i}mage \nj{q}uality and \nj{this strategy of training LR to mimic new HR can lead to a} more efficient SR network. \nj{Up to a certain point, by repeating this process multiple times, better and better images are obtained}. This recurrent leaning \nj{strategy} for SR can be a good solution for downsizing convolution networks and \nj{making a more efficient SR network}. \nj{To measure the enhanced image quality, for the first time in this area of super-resolution and image enhancement, we use VIQET \cite{VIQET} MOS score which reflects human visual quality more accurately than the conventional MSE measure.} 
    \end{abstract}


\section{Introduction}

\nj{Nowadays}, 2K (1920$\times$1080) videos are \nj{widely used in} digital broadband systems. From the viewpoint of image quality, there \nj{are} two major \nj{sources of impaired quality in the current video images}. The first \nj{one} is \nj{the intrinsic} limitation of image capture systems such as lens optics, sensor resolution, focusing performances and ISO noise. These defects make \nj{inherent limitations in image quality}\sam{. Therefore} almost every videos cannot show \sam{ideal 2K image quality, with maximum frequencies.} 
\nj{The} second \nj{source of corruptions is due from} image processing such as compression, resolution converting and noise reduction. \nj{Most components of a broadband system have a fixed resolution and they require} unavoidable needs of image scaling. \nj{Particularly, in order to transmit sources with lower resolution made in the past via a broadband system, video upscaling is inevitable.} Also \nj{most} of contents providers 
\nj{use} heavy compression 
\sam{and restoring the lost information is a difficult task \nj{which requires enhanced} techniques.}

\begin{figure}
\begin{center}
   \includegraphics[width=1\linewidth]{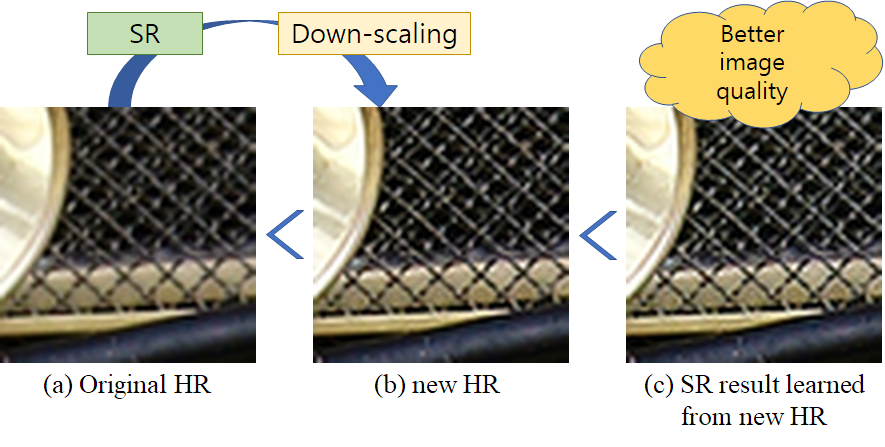}
\end{center}
\vspace{-5mm}
   \caption{\textbf{\nj{Proposed} Concept.} It would be better to learn through super-resolution-processed images. \nj{The figures are} $HR_0$ (original), $HR_1$ and $HR_3$ \nj{obtained by our method}. With the same network, \nj{(c) shows better quality than} (b). }
\vspace{-2mm}
\label{fig:concept}
\end{figure}

\nj{The early image enhancement techniques were mostly} at the level of unsharp-masking, which is \nj{just} a simple amplification \nj{operation}. With the advent of high-resolution panels, super-resolution \nj{techniques have been} developed to improve the original resolution. \nj{In} recent years, machine learning \nj{technologies are widely used} and deep learning has \nj{achieved state-of-the-art} super-resolution \nj{performance}. However, \nj{as} the resolution of the image \nj{as well as} the complexity of the algorithm increases, \nj{implementation becomes harder and harder. } 
\nj{Therefore, it is necessary to develop} cheaper and more effective algorithms \nj{for image enhancement and super-resolution}.

\nj{In this paper, as shown in Figure \ref{fig:concept}, we introduce} the concept of iterative learning as an approach for image enhancement. In general, if the same \nj{algorithm} is applied repeatedly, the result will be reinforced. Especially in deep learning, we can approach \nj{this idea} without \nj{additional computational cost at inference time. What we need is additional training} with \sam{the new HR, made in the previous stage. Through this recurrent learning, the image converges to a better quality, and it \nj{is} possible to obtain \nj{an enhancement} over the existing super-resolution with a smaller-sized network. \nj{By training a small network with 200K parameters three repetitions, we could get images with better quality compared to a model with over 1M parameters.} } 


\nj{The contributions of this paper are as follows:}

1. By showing an effective \nj{way of repeatedly training a simple feed-forward network for super-resolution} 
through recurrent learning, \nj{we improve} the efficiency of super-resolution \nj{achieving} similar results to those \nj{of complex deep networks} 
without \nj{much computational} cost. 

2.	By shifting the evaluation criteria of the image quality from a simple \nj{arithmetic measure} like Mean-Squared-Error \nj{(MSE)} to a user-centered measure of \nj{VIQET \cite{VIQET}} Mean-Opinion-Score \nj{(MOS)}, \nj{which can be computed in an automated way without human involvement,} we can focus on additional \nj{subtle image quality} improvement \nj{and set the number of repetitions appropriately}.




\section{Related work}
\subsection{\nj{Image enhancement}}
Along with the development of computer vision and the spread of digital broadcasting, image processing techniques have been widely used in image production and reproduction. \nj{These techniques} can be roughly divided into \nj{noise} reduction and \nj{image} enhancement. The former includes technologies such as de-noising and de-interlacing that remove artifacts that should not be present in the original image. In the latter, sharpness and contrast \nj{are} used to amplify \nj{some} components in the original image \nj{for better image quality}. These algorithms typically perform a series of operations that widen the min-max range of the original \nj{image}, such as \nj{the} Laplacian filtering. The Laplacian filtering is inexpensive and effective, but it is not possible to perform \nj{more} complex processing \nj{such as restoring} missing texture components or broken edges. 

There are several image enhancement algorithms that \nj{resolve} these weaknesses. \nj{The first one is} improving the adaptiveness \nj{of the processing region} such as \nj{Adaptive Unsharp Masking \cite{826787}, which selects the processing region wisely}. The second \nj{one} is to use two or more weights in the function to enable more complex processing, which includes \sam{Bilateral Filter \cite{4476197} and NLM Filter\cite{1467423}}. There \nj{are huge amount of} related works, but with the advent of super-resolution, research on conventional image enhancement has diminished.
{This is \nj{mainly} because the improvement is due \nj{from} the amplification of \nj{some components}, and the original frequency is maintained. Ultimately, it requires an approach such as super-resolution in order to change the shape of the original. We present a direction for developing new super-resolution \nj{images} for better \nj{quality}.}

\subsection{\nj{Super-resolution}}
Before the boom of deep learning, as a better upscaler, \nj{super-resolution algorithms were} developed. \nj{These techniques have mainly been} applied to a \nj{display} system such as a television. \nj{It} is proposed as a solution to compensate for \nj{the mismatch between the input image size and the panel resolution}. 
There are two \nj{main} categories of \nj{this} technology: \nj{1)} utilizing machine learning \nj{algorithms} and \nj{2) weighted image blending using the patches} from the original \nj{image}. The former is similar to the super-resolution \nj{approaches} through the current deep learning, but \nj{with simple architectures} consisting of only one layer. The latter is a self-image generation method that retrieves the most similar patches \nj{from the surrounding areas of images after upscaling} \cite{5459271}. 
Conventional super-resolution \nj{methods are} well documented in \nj{\cite{5466111}}. \nj{They} can perform upscaling while keeping the sharpness of an edge. However, there are technical limitations such as \nj{the problem for} a large upscaling ratio or impossibility of \nj{restoring} complex textures.  {\nj{These require} huge computations, but it was difficult to learn enough with a human-designed shallow network. Recently, \nj{deep neural networks have} been improved to enable more powerful super-resolution processing. This is why we use deep learning as a solution.}

\subsection{Super-resolution with deep learning}
\nj{Recently, as} a branch of deep learning \nj{research}, `single image super-resolution' has \nj{emerged}. It makes down-sized \nj{low-resolution (LR)} images and \nj{trains} convolution networks \nj{so as to reproduce the original images from the LR images}. 
\nj{Compared to the revolutionarily simple} SRCNN \cite{10.1007/978-3-319-10593-2_13}, \nj{the architecture has become} heavier and more complex. 
VDSR \cite{Kim_2016_CVPR} introduced VGG and ResNet into \nj{the} super-resolution filed \nj{in} 2015. ESPCN \cite{Shi_2016_CVPR} \nj{and} DRCN \cite{Kim_2016_CVPR_DRCN} present various aspects of residual structure and \nj{the} view of upscaling \nj{in} 2016. SRGAN \cite{Ledig_2017_CVPR}, SRResNET \cite{SRResnet}, DRRN \cite{Tai_2017_CVPR}, EDSR \cite{Lim_2017_CVPR_Workshops}, DenseSR \cite{Tong_2017_ICCV}, MDSR \cite{Lim_2017_CVPR_Workshops} \nj{and} Memnet \cite{Tai_2017_ICCV} \nj{developed} Resnet architecture and \nj{adjusted the} GAN solution into super-resolution \nj{in} 2017. In 2018, super-resolution is now more widely used in terms of utilization. 
A good example is  \cite{Yoo_2018_CVPR}. \nj{There is a research utilizing a recursive method for super-resolution \cite{Dahl_2017_ICCV}. }
\nj{Different from ours, 
it tries to} \sam{produce \nj{a} face image from \nj{an} 8$\times$8 tiled mosaic }
through repeated application of super-resolution, \sam{\nj{which is} similar to info-GAN \cite{NIPS2016_6399} \nj{that} produces a facial image.}

\sam{Regardless of the super-resolution algorithm,} 
\nj{the general operation principle is similar. Basically, the} single image super-resolution is learned by back propagation so as to \nj{reduce} the difference between the \nj{original and the generated image through the SR network from an LR image}. 

There are some limitations in these works. First, in the low-resolution images used in other SR papers, the improvement of the image \nj{quality} by \nj{increasing the resolution} is remarkable. However in a high-resolution image, the improvement caused by an increase in resolution \nj{is} not obvious. Second, the advances in super-resolution have led \nj{to an increased performance sacrificing the computational complexity.} 
\nj{This high computational demand makes it hard to implement super-resolution with hardware. Therefore,} it is necessary to develop a method that can achieve better performance with a more simple network. Finally, \nj{since current super-resolution algorithms are} learned \nj{only from the} original \nj{images, they target} the original and \nj{we cannot expect} image quality beyond the original. {\nj{In the following section, we} present a new approach to get \nj{images with better quality at a low cost} through recurrent learning.}

\section{\nj{Proposed method}}

\subsection{\nj{Recurrent Training Strategy (RTS)}}
\begin{figure}
\includegraphics[width=\linewidth]{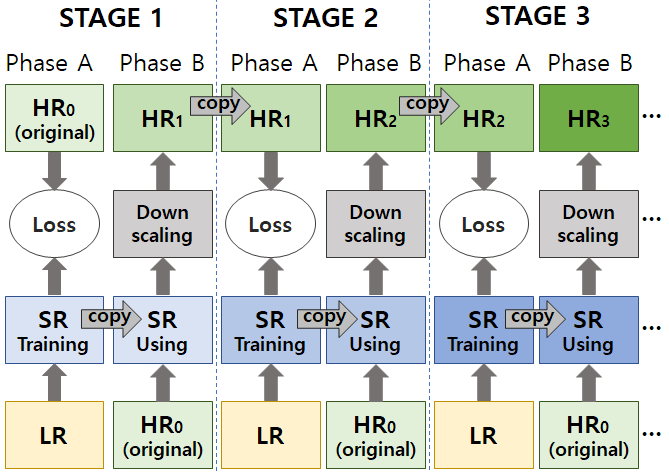}
\vspace{-4mm}
\caption{\nj{\textbf{Recurrent Training Strategy(RTS).} LR is a low resolution image obtained by down-scaling the original image ($HR_0$). One stage is composed of the SR training phase (Phase A) and the image enhancement phase (Phase B). By successive application of these two phases, we can obtain a better SR network and enhanced images.} \sam{Although it takes multiple times to learn, \nj{at inference time,} it is possible to obtain \nj{images with quality equivalent to the finally learned $HR_{n}$ level} at once \nj{by feed-forwarding with the learned weights}. 
} }
\label{fig:overall}
\end{figure}

The core of this paper is the recurrent \nj{training strategy of an SR network which is achieved by} reducing the \nj{size of the output image of the SR network} to its original size and \nj{retraining the SR network with this new target}. The \nj{proposed} recurrent \nj{training strategy} is \nj{briefly} shown in Figure \ref{fig:overall}.

\nj{As can be seen in the figure, a stage of RTS} consists of two phases. In the first phase \nj{(Phase A), SR network is trained}. \sam{In this phase, the SR is learned} so that the target image is obtained from the degraded LR image, and the weight of the super-resolution network is updated. The LR image is a \nj{low quality} image which is corrupted through down-scaling, adjusting compression \nj{and so on.} 
In Stage 1\sam{, the target is set to the original image, $HR_0$ and} the network is \nj{tuned such} that the LR is improved to the image quality of the original $HR_0$ level. 

The second phase, Phase B, is the process of creating a new target image using the network \nj{obtained in} Phase A. \nj{This is done by inputting the original image $HR_0$ to the trained network} of Phase A \nj{and we can expect that the output image $SR_1$ has higher quality than the original image $HR_0$. By downscaling the output image $SR_1$, we can obtain a new image $HR_1$ which acts as a target in the next stage. }  
Generally, super-resolution is accompanied by increasing the size of the image to increase the image resolution. So the additional down-scaling process is required because the resolution must be the same as the original for \nj{retraining the network}. 

To summarize, we have to go through two steps to get a new HR \nj{image}. The former is the process of obtaining the super-resolution \nj{function $SR(\cdot)$}, and the latter is the \nj{process of obtaining enhanced image by down-scaling the output image of the SR network back to the original size}. Then, we perform this process repeatedly on a stage basis, so that the same super-resolution processing can be \nj{more effective}. 
\nj{The overall procedure of Stage $n$ can be summarized as follows:
\begin{equation}
    \begin{split}
        &\text{Phase A: } SR_n = \argmin_{SR} \frac{1}{2}||HR_{n-1} - SR(LR)||^2 \\
        &\text{Phase B: } HR_n = d(SR_n(HR_0)).
    \end{split}
\end{equation}
Here, $SR(\cdot)$ and $d(\cdot)$ denote the super-resolution network and the down-scaling operation, respectively. $LR$ is an image which has lower resolution than the original $HR_0$. 
}

By applying super-resolution to the original image, a \nj{higher quality image is expected to be}  obtained. Even if the obtained image is down-scaled to the original size, \nj{components from the additionally} generated resolution \nj{may remain}, this is the same \nj{as an UHD (Untra HD) channel that looks better than the existing channels on a FHD \nj{(full HD)} TV}. 
If \nj{$HR_n$} has a better image quality than \nj{$HR_{n-1}$}, we can assume that the super-resolution \nj{network} can learn from \nj{this} better target, and after \nj{the learning, the resultant network will be better.} 
\nj{Therefore, assuming} that we go through this additional process in a direction that gradually improves the results, we will eventually reach the maximum improvement point of each image. And it \nj{will perform better than} the \nj{current} super-resolution \nj{network trained to mimic the original image}.

\begin{figure*}
    \includegraphics[width=0.95\linewidth]{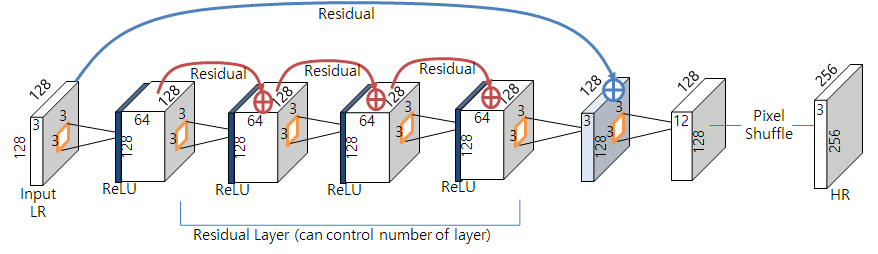}
    \caption{\textbf{Used network for recurrent learning.} {This network was developed exclusively for recursive learning systems. The biggest difference from the existing SR network is \nj{that it was} designed to produce the best result at the \sam{blue layer} \nj{where the} residual is added \nj{with only 3 channels.}} 
    \nj{The model} is also very light and easy to control the number of layers.}
\vspace{-2mm}
\label{fig:network}
\end{figure*}

\subsection{\nj{Reason for enhanced quality}}
If $y$ is the original image, $U$ is the up-scaler, $D$ is the down-scaler, and $R$ is the residual component, the super resolution trained by \nj{$x=D(y)$} can be expressed as follows: 
\begin{equation}
    SR(x) = U(x) + R(x),
    \label{eq:srx}
\end{equation}
\begin{equation}
    SR(D(y)) = U(D(y)) + R(D(y)).
\end{equation}
Here, we can suppose $D(U(y))$ is \nj{the} same as $y$, and $U(D(y))$ \nj{becomes} \nj{a locally blurred version of} $y$: 
\begin{equation}
    D(U(y)) = y, \quad
    U(D(y)) = \bar{y} 
    \label{eq:udy}
\end{equation}
\nj{Combining (\ref{eq:srx}) and (\ref{eq:udy}), we get }
\begin{equation}
    R(x) = SR(x) - \bar{y}.
\end{equation}
\nj{Because} the goal of learning is to minimize \nj{the error $||y-SR(x)||$}:
\begin{equation}
    SR^* = \argmin_{SR} \left( \frac{1}{2} \left \| y - SR(x)\right \| ^{2} \right),
\label{eq_3}
\end{equation}
\nj{if} enough learning has been done, $R(x)$ \nj{becomes} similar to $y-\bar{y}$:
\begin{equation}
    R(x) \approx y - \bar{y}.
\label{eq:approx1}
\end{equation}
In the first stage, after \nj{the} super resolution is fully learned, \nj{instead of $D(y)$,} we can input $y$ to \nj{the SR network}, \nj{then it becomes}
\begin{equation}
    SR_{1}(y) = U(y) + R_{1}(y).
\end{equation}
\nj{By down-scaling this, we get the target image for stage 2 which can be expressed as}
\begin{equation}
    HR_{1}(y) = D(SR_{1}(y)) = D(U(y) + R_{1}(y)). 
\end{equation}
\nj{Assuming a linear\footnote{For a nonlinear down-scaler such as nearest neighbor or bicubic interpolation, we can take Taylor series expansion.} down-scaler, as $D(U(y)) = y$, it becomes} 
\begin{equation}
    HR_{1}(y) = y + D(R_{1}(y)).
\end{equation}
\nj{The distributions of zero-mean residual components of different sized images will be approximately equal, that is $p(z|z\in R_1(x)) = p(z|z\in R_1(y))$, unless there are not much high frequency components only existing in $y$ and not in $x$. By down-scaling $R_1(y)$, the distribution will be more contracted near zero and we can approximate $D(R_1(y))$ with a contracted version of $R_1(x)$ as }
\begin{equation}
    D(R_{1}(y)) \approx \alpha R_{1}(x) , \quad ( 0 \le \alpha \le 1 ).
\label{eq:Dtoalpha}
\end{equation}
Finally, with (\ref{eq:approx1}), $HR_{1}(y)$ \nj{can} be approximated \nj{as}
\begin{equation}
    HR_{1}(y) \approx y + \alpha R_{1}(x) \approx y + \alpha (y-\bar{y}).
\end{equation}
\nj{The above equation} has high similarity \nj{with} unsharp masking. In other words, $HR_{1}(y)$ has a high probability that the \nj{components with a large deviation from the mean} are amplified. And it can be expected that \nj{the new target image} $HR_{1}(y)$ can \nj{have} improved sharpness compared to $y$.

When \nj{the} super resolution \nj{network is trained with this new target, it becomes 
\begin{equation}
    HR_2(y) \approx y + \alpha R_2(x) \approx y + \alpha(y-\bar{y}) + \alpha^2 (y-\bar{y}).
\end{equation}
This is obtained by replacing $y$ in (\ref{eq:approx1}) with the new target $y+\alpha(y-\bar{y})$ in computing $R_2(x)$.
As stage goes on, it becomes} 
\nj{
\begin{equation}
    HR_{n}(y) \approx y + \sum_{i=1}^n \alpha^i (y-\bar{y}). 
\label{final_eq}
\end{equation}
}
\nj{Since $\alpha$} is smaller than 1, \wacvadded{(Experimental results show that the average $\alpha$ for stage 1 on DIV2K valid set is 0.779 with the standard deviation of 0.098.)} \nj{higher order $\alpha$ terms} will disappear and $HR_n$ can be expected \nj{to} convergence to \nj{a certain point $\hat{y}$ not far from $y$}. 
\nj{Using the SR network, the actual difference $y-\bar{y}$ is learned by a nonlinear function \wacvadded{and it makes hard to guarantee convergence, these non-linearity} has much better performance than the simple USM algorithm. As iteration goes on, the difference tend to converges to $\hat{y}-\bar{y}$ and the image quality of $\hat{y}$ becomes better than that of the original image $y$.}

\subsection{\nj{Configuration of super-resolution network}}
We set up a convolution network based on the residual layer. The use of multiple layers of residual type has the advantage of automatically adapting image filters of various sizes. In order to make use of these advantages and to easily control the number of parameters, we constructed the \nj{network shown in Fig. \ref{fig:network}}. 

{This network is designed exclusively for recurrent learning, which uses \nj{a subsequent} down-scaler \nj{after the operation of super-resolution}. The biggest difference from \nj{the} regular super resolution network \nj{is that} it produces the best output of the same resolution with original, not \nj{in an} upscaled domain. Learning proceeds in the same way as the existing super resolution, but since the expected output is synthesized in 3ch at the original resolution, optimum result is produced at the corresponding step.}

It is composed \nj{of an} input module, \nj{an} output module and three \nj{64-channel residual networks using 3$\times$3 filters} 
The input module \nj{consists of} 128$\times$128 RGB (3ch) input and 64ch output. After the input module, a \nj{series of residual layers, all of which have the same number of channels (64 ch) with 3$\times$3 convolutions, is applied. As shown in the figure,} the base model uses 3 layers. In the output module, the final residual \nj{layer does not make use of} ReLU because \nj{the difference between the input and the output} will have both \nj{a positive and negative} value. {Here, the output with blue residual will be next stage's HR, because the next channel extension and pixel shuffle has no ReLU and it can be canceled by the down-scaler.} The residual layer is \nj{based} on SRResNet \cite{SRResnet} and VDSR \cite{Kim_2016_CVPR}. \nj{The final residual without ReLU is} from VDSR, and \nj{the} inner residual layer \nj{shares the} concept \nj{with} SRResNet.

Finally, we need a down-scaler to convert the \nj{up-scaled} image to the same size as the original, using the Lanczos scaler. Lanczos filter has better performance than linear \nj{and bicubic filters}, and it makes $\alpha$ greater in \nj{(\ref{eq:Dtoalpha})}.
%

\begin{table}
\caption{\textbf{Image difference ratio of validation set.} \nj{In our model with the proposed} recurrent \nj{training strategy}, the change of difference ratio \nj{from the previous stage is minimum at stage 4.}
}
\begin{center}
\begin{tabular}{|c|c|c|}
\hline
	Stage N&Difference Ratio &Delta from (N-1)\\
	& form the Original(\%)& \\
\hline\hline
	\texttt{HR1}&0.95&0.95\\
	\texttt{HR2}&1.35&0.40\\
	\texttt{HR3}&1.47&0.12\\
	\texttt{HR4}&1.48&\textbf{0.01}\\
	\texttt{HR5}&1.51&0.03\\
	\texttt{HR6}&1.55&0.04\\
	\texttt{HR7}&1.62&0.07\\
	\texttt{HR8}&1.76&0.14\\
\hline
\end{tabular}
\end{center}
\vspace{-2mm}
\label{table:image_diff}
\end{table}

\begin{figure}
\centering
\includegraphics[width=0.8\linewidth]{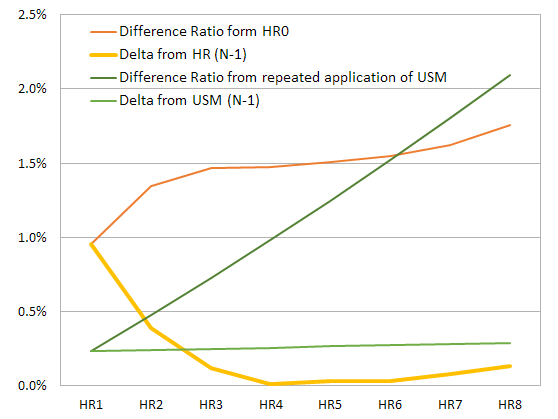}
\caption{\textbf{Change of image difference \nj{vs.} stage.} It can be seen that the difference converges to almost zero in HR4 (Yellow Line). But with repeated application of unsharp masking, we can observe the delta \nj{increases} linearly (Green Line). }
\label{fig:image_diff_graph}
\end{figure}

\begin{figure*}
\begin{center}
    \begin{tabular}{c}
    \hspace{-5mm}(Original)\hspace{23mm}(HR1)\hspace{27mm}(HR2)\hspace{27mm}(HR3)\hspace{27mm}(HR4)\\
    \vspace{-4mm}
    \\
    \hspace{-3mm}
    \includegraphics[width=1.35in]{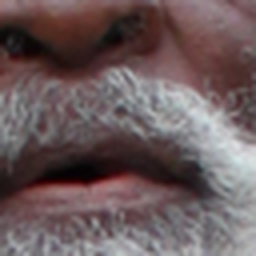}
    \includegraphics[width=1.35in]{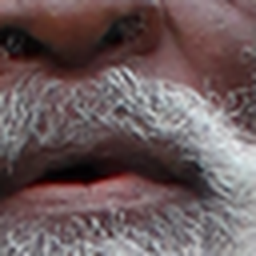}
    \includegraphics[width=1.35in]{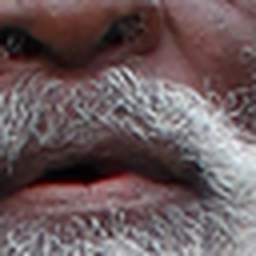}
    \includegraphics[width=1.35in]{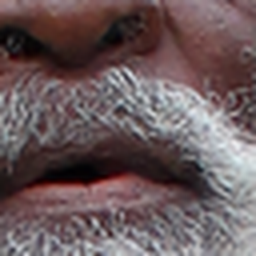}
    \includegraphics[width=1.35in]{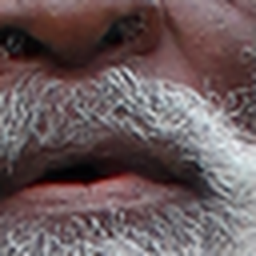}\\
    \vspace{-4mm}
    \\
    \hspace{-3mm}
    \includegraphics[width=1.35in]{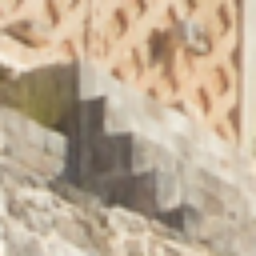}
    \includegraphics[width=1.35in]{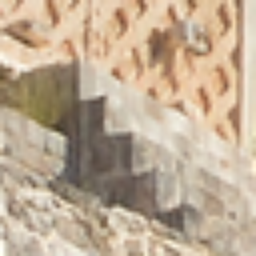}
    \includegraphics[width=1.35in]{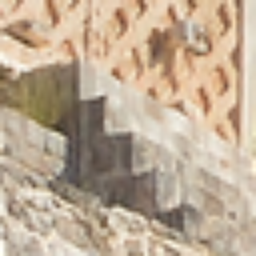}
    \includegraphics[width=1.35in]{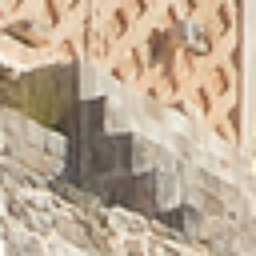}
    \includegraphics[width=1.35in]{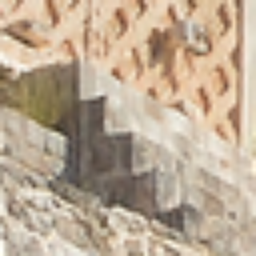}\\
    \vspace{-4mm}
    \\
    \hspace{-3mm}
    \includegraphics[width=1.35in]{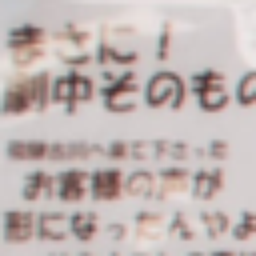}
    \includegraphics[width=1.35in]{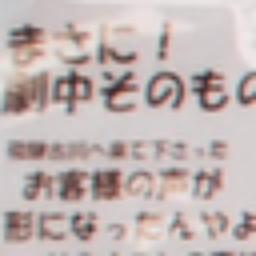}
    \includegraphics[width=1.35in]{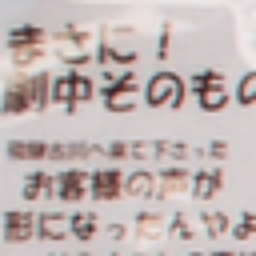}
    \includegraphics[width=1.35in]{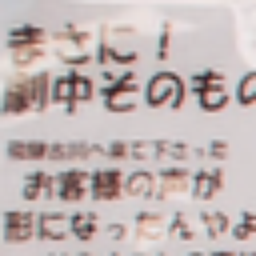}
    \includegraphics[width=1.35in]{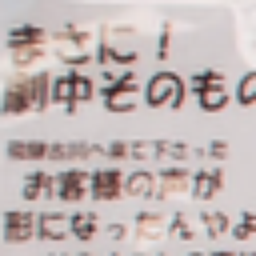}\\
    \vspace{-4mm}
    \\
    \hspace{-3mm}
    \includegraphics[width=1.35in]{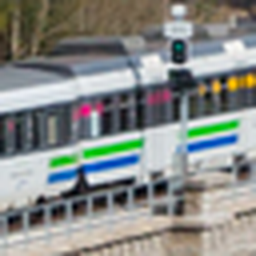}
    \includegraphics[width=1.35in]{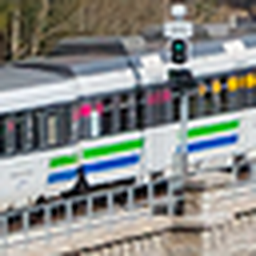}
    \includegraphics[width=1.35in]{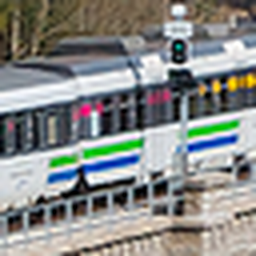}
    \includegraphics[width=1.35in]{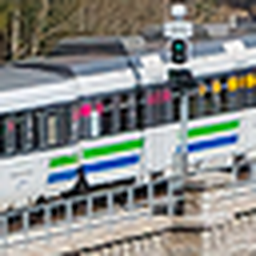}
    \includegraphics[width=1.35in]{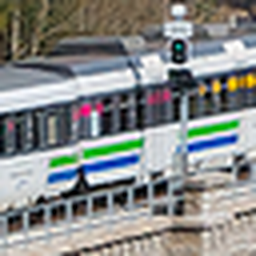}\\
    \\
    \end{tabular}
\vspace{-5mm}

    \caption{\textbf{Comparison of \nj{output} images \nj{in} each stage.} \nj{For better comparison, each row shows a part \wacvadded{(64 $\times$ 64 pixels)} of an image} in DIV2K image set. \wacvadded{As described in the introduction, the originals are naturally blurred in the capturing and the transmission process.} The stage number increases \nj{from left to right}.  \nj{As the stage} increases, it can be observed that \nj{the texture is clearly seen and} the result becomes clearer. {It is easily identifiable in a man's beard, in the particle of the bricks, and in the clarity of the letters. }}
\label{result_in_image}
\vspace{-3mm}
\end{center}
\end{figure*}

\subsection{Setting the number of stages}
In our proposal, \nj{the number of stages $N$ can be set as the point where the quality of $HR_{N+1}$ is worse than that of $HR_{N}$ and the corresponding network $SR_N$ can be used as the final super resolution network.} 
The important thing here is how to judge \nj{whether the output image} gets better or not. As a result of \nj{our} experiment, \nj{we can consider a couple of} solutions. The easiest approach is to measure the quality of the actual \nj{validation} set to see if this score is increased. This is intuitive, but it has the disadvantage that it is difficult to use it because of the time and effort required for measurement. The second method is to confirm whether the change \nj{$||HR_{N+1} - HR_{N}||$} becomes larger or smaller as the repetition progresses. As the output of each stage approaches a certain ideal point, the difference \nj{will} gradually decrease. When the output of the stage deviates from the ideal point, the divergence starts and the difference \nj{will increase}. This will be explained in more detail in the \nj{following experimental results}.



\section{Experimental result}

\subsection{Training setup}
\nj{In our experiment, we used the DIV2K image dataset \cite{DIV2K}, which contains 2K RGB images. }
This is to prove \nj{the performance of image enhancement on contents in the FHD level which} occupies the majority of \nj{current} broadcasting \nj{contents}. The DIV-2K \nj{dataset} consists of 800 train sets and 100 validation sets. \nj{From the 800 DIV-2K training set, we made our \nj{training} database which was} constructed by ripping 100 random patches \nj{with a} 256$\times$256 size. After making \nj{the original images, $HR_0$}, we adjusted the JPEG compression and \nj{down-scaled the patches} to \nj{the} half size to create the LR images. \sam{The learning was carried out from the 80,000 pairs made in this way.}

\nj{
In all of our experiments, we used MSE as a loss function and Adam as an optimizer. Batch size of 8 was used with learning rate of 0.001. We trained the network for 20 epochs.
}

\subsection{Convergence of recurrent learning}
In order to confirm whether the \nj{proposed} recurrent learning \nj{converges} to the maximum improvement point or simply diverges as if the Laplacian filter is applied multiple times, we observed the \nj{difference} of \nj{output images between consecutive stages}. The \nj{image difference} is an objective indicator that shows the average difference of all the pixels in \nj{an} image in \nj{terms of ratio}. It can be similar to MSE, but it is normalized to the image's brightness and resolution. \nj{It is a more interpretable measure in that it is} more intuitive \nj{telling the percentage of the image}  changed. The average image difference of the \nj{100 validation images} in each stage \nj{is shown in} Table \ref{table:image_diff}.

\nj{Figure \ref{fig:image_diff_graph}} shows Table \ref{table:image_diff} to a graph. It shows that the learning converges to a certain point. And it can be seen that the \nj{change of difference starts to increase} after a specific point, the Stage 4.
In order to confirm that the recurrent learning is different from the repetitive application of \nj{a} simple filter, we \nj{repeatedly applied unsharp-masking (USM) filter to the original image, which is shown by} the green line. The difference \nj{increases} linearly and \nj{image difference} \nj{from Stage (N-1) does not converge ever}. Also, from the \nj{observation that the} image difference is linearly increased \nj{by repeated USM application}, we can confirm that the image difference is properly measured.

\begin{table*}
\caption{\textbf{Result of VIQET image quality measurement.} The final MOS score in this table is obtained based on the detailed score of 10 items excluding \textsc{Flat Region Index} which results in NaN for all the cases. 
Except for over and under exposed ratio, a high score lead to a high MOS, which means a better quality. }
\begin{center}
\small
\begin{tabular}{|c|c|c|c|c|c|c|c|c|c|c|c|}
\hline
	&MOS&Resol&Edge&SN&Texture&Satu&Color&Illumi&Dynamic&Over&Under\\
	Name&Score&ution&Acutance&Index&Acutance&ration&Warmth&nation&Range&Expose&Expose\\
\hline\hline
    \texttt{LR}&	3.645&	2.8&	52.6&	311.9&	96.3&	116.8&	107.7&	156.8&	101.0&	0.1&	0.4\\
    \texttt{HR0}&	3.825&	2.8&	56.6&	314.8&	108.7&	117.3&	107.8&	163.6&	101.0&	0.1&	0.4\\
    \texttt{HR1}&	3.962&	2.8&	82.7&	322.3&	150.7&	117.7&	107.8&	182.7&	101.3&	0.1&	0.4\\
    \texttt{HR2}&	4.028&	2.8&	92.5&	325.3&	167.8&	\textbf{118.4}&	\textbf{108.0}&	190.0&	101.2&	0.0&	0.4\\
    \texttt{HR3}&\textbf{4.057}&2.8&\textbf{94.3}&	326.2&	\textbf{171.8}&	118.0&	107.8&	\textbf{191.4}&	101.2&	0.0&	0.4\\
    \texttt{HR4}&	4.050&	2.8&	93.1&	326.7&	171.4&	117.1&	107.2&	191.1&	101.3&	0.1&	0.4\\
    \texttt{HR5}&	4.048&	2.8&	92.1&	326.7&	170.5&	117.7&	107.6&	190.7&	101.3&	0.1&	0.4\\
    \texttt{HR6}&	4.038&	2.8&	89.5&	327.6&	168.1&	116.0&	107.8&	188.2&	101.1&	0.0&	0.4\\
    \texttt{HR7}&	4.030&	2.8&	87.9&	329.0&	166.9&	114.6&	107.3&	186.8&	101.1&	0.1&	0.4\\
    \texttt{HR8}&	4.030&	2.8&	87.3&	\textbf{331.1}&	166.1&	111.7&	106.1&	186.2&	101.2&	0.1&	0.4\\
\hline
\end{tabular}
\end{center}
\vspace{-5mm}
\label{table:mos}
\end{table*}

\subsection {The result images}
\nj{Figure \ref{result_in_image}} shows \nj{some samples of the output images in} each stage. It shows how the \nj{resultant images} for each stages are being improved. All images are down-scaled with the same resolution as the original, and we can visually confirm that the \nj{quality} of the images \nj{is} improved. The improvement level at each stage is similar to the slope of the graph in \nj{Fig. \ref{fig:image_diff_graph}}. \nj{As stage goes on}, it can be observed that the texture is clearly seen and the result becomes clearer. 
{It can be said that recurrent learning converges to a certain level of improvement, since there is little difference between $HR_{3}$ and $HR_{4}$ \nj{in all the samples}.}

\begin{figure}
\centering
\includegraphics[width=0.8\linewidth]{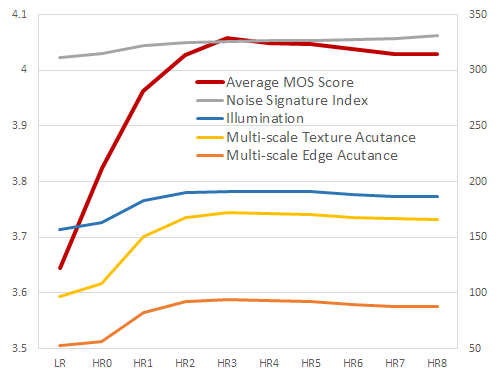}
\caption{\textbf{The MOS results for each stage.} The final score, MOS (Red line), peaked at $HR_{3}$. {The increase in MOS is similar to the convergence of the difference, yellow line in Fig.\ref{fig:image_diff_graph}.
Also the \nj{specific measures in other colors} tend to be increasing until $HR_{3}$ and then \nj{they saturate} or decrease slightly.}
}
\vspace{-3mm}
\label{fig:MOS}
\end{figure}

\subsection {Verification using VIQET tool}
In order to clearly demonstrate the effectiveness of image enhancement, we looked for ways to more objectively measure the image quality. Many non-reference quality measure models \nj{exists}, but most \nj{of them need} pre-learning the references and \nj{check} how close the measured image is from \nj{the} learned. This \nj{is} not what we want, because we \nj{do not} have any reference. Fortunately, the \textit{Video Quality Experts Group} (VQEG) has developed a tool that measures the image quality and \nj{developed} the Mean Opinion Score (MOS) of \nj{an} image as 0 (Poor) to 4.5 (Best). Using this tool, we can evaluate the image quality improvement for each stage objectively.

\nj{In the} VQEG image quality Evaluation Tool (VIQET) \cite{VIQET}, the average MOS score is derived for each image through a total of 11 measures: 
{\small 
\textsc{1. Resolution, 2. Multi-scale Edge Acutance, 3. Noise Signature Index, 4. Flat Region Index, 5. Multi-scale Texture Acutance, 6. Saturation, 7. Color Warmth, 8. Illumination, 9. Dynamic Range, 10. Over Exposed ratio, 11. Under Exposed ratio}}. \nj{To use VIQET, we must set a category of an image among 4 categories: 
{\small
\textsc{Outdoor Day - Landscape, Indoor - Wall Hanging, Indoor - Arrangements}
} and 
{\small \textsc{Outdoor Night - Landmark}}. 
Most of the DIV-2K images \nj{are} bright images taken from the outside and we simply set all \nj{the images} to the \textsc{outdoor day} category.} The \nj{detailed average} scores of 100 DIV-2K \nj{validation} images for each stage are shown in Table \ref{table:mos}. 

The key point is that the overall score, MOS value, increases with \sam{the stage}. The result proves that ``The image quality can be increased by recurrent learning''. The MOS results for each stage are shown in Fig. \ref{fig:MOS}. Based on the MOS, it can be confirmed that the image quality from the original \nj{3.825} was improved up to 4.057. The increase is 0.232 points \nj{from the original}, which is larger than the difference between LR and \nj{the} original, 0.180. \sam{It prove} that image enhancement through recurrent learning is efficient. \nj{There} is a limitation \nj{in improving the image quality based on the difference between $HR_0$ and LR with one-time learning and we have resolved this by providing a new target obtained through recurrently training the network.} 
As can be seen in Fig. \ref{fig:MOS}, the MOS increase is largely due to the improvement of four terms. The first one is the improvement of multi-scale edge acutance \nj{value.} It means the edge got to be more sharp. The second is the Noise Signature Index, which means that the distinction between image and noise is more clear. The \nj{third} one is multi-scale texture acutance, and it means increment of {the level of activity and detail in the scene. Lastly, illumination means the light has become more enriched, also \nj{indicating} that the image has more detail.}

\begin{table}
\caption{\textbf{MOS results for various networks.} This table shows that there is an additional improvement in the index after HR1. It means the application of Recurrent Training Strategy (RTS) is meaningful regardless of the network. And we can confirm the proposed network is more suitable for recurrent learning. \nj{Images are available in the supplementary material.}}
\begin{center}
\small
\vspace{-3mm}
\begin{tabular}{|c||c|c|c|c|c|}
\hline
	Network&Param&HR0&HR1&HR2&HR3\\
\hline\hline
    \texttt{\textbf{Proposal}}&&&&&\\
	\texttt{Residual1}  & 41K   &2.384&2.639&2.701&2.712\\
	\texttt{Residual3}  & 114K  &2.384&2.637&2.719&2.720\\
	\texttt{Residual6}  & 225K&2.384&\textbf{2.644}&\textbf{2.725}&\textbf{2.744}\\
\hline
    \texttt{\textbf{Reference}}&&&&&\\
    \texttt{SRCNN}      &57K    &2.384&2.603&2.680&2.680\\
    \texttt{VDSR}       &666K   &2.384&2.599&2.668&2.669\\
    \texttt{SRResNet}   &1.5M   &2.384&2.617&2.720&2.703\\
    \texttt{EDSR(64$\times$16)}&1.7M   & 2.384 &2.632&2.725&2.737\\
\hline
\end{tabular}
\end{center}
\vspace{-7mm}
\label{table:comparison}
\end{table}

\subsection{RTS \nj{on} various networks}

\nj{To investigate the characteristics of the proposed RTS more clearly and to check whether the proposed RTS} is still effective with other \nj{algorithms}, we test \nj{RTS on} several other networks 
as well as the proposed network.
\nj{RTS was applied to total} six networks \nj{and the performances are shown in Table \ref{table:comparison}}. \nj{In the table, the above three networks are designed based on our proposal shown in Fig. \ref{fig:network} by changing the number of residual layers in the middle to 1, 3 and 6, respectively.} 
The proposed architecture in Fig. \ref{fig:network} has the advantage of being able to easily compare the number of parameters in the network \nj{considering the number of repetitions}.  For example, a comparison between learning once with doubled parameters and learning twice with halved parameters \nj{is possible}. In the three bottom rows,  we have \nj{chosen} SRCNN, VSDR, \wacvadded{SRResNet and ESDR} to check whether our \nj{recurrent training strategy} is meaningful even in \nj{a} heavier super-resolution network. \nj{By this experiment, the characteristics} of \nj{the proposed RTS on various networks can be checked} from the simplest network, SRCNN, to \wacvadded{a} latest heavy network, \wacvadded{EDSR}.

Unfortunately, the existing networks are so heavy that the GPU memory \nj{overflows}, making it impossible to keep the original size of a DIV2K image intact. So we have to divide each image into 4$\times$2 and processed \nj{them independently}. The MOS result is \nj{the} averaged value of 800 pieces of divided 100 \nj{validation images}. The segmentation of the image caused a decrease in the content and resolution, which resulted in a decrease in the MOS score. However, since this segmentation is applied equally to all the six \nj{networks}, the tendency is comparable and the experiment is fair. Table \ref{table:comparison} shows the resultant MOS scores as well as the number of parameters in each network. 

In \nj{Table} \ref{table:comparison}, the scores of HR2 and HR3 are higher than \nj{HR1 at all times}. This suggests that recurrent learning may lead to additional performance improvement in existing algorithms. In addition, it is shown that the improvement \nj{due from} the learning method is more meaningful than the improvement \nj{caused by using a different} network. It means that the \nj{quality of an image} is more dependent on the number of repetitions in RTS than the size of the network, and as a result, the proposed RTS can produce good results even with light networks.
%
Although the \nj{amount of increase in} MOS is different for each algorithm, the tendency of learning is similar. From this, we can say that applying \nj{RTS} makes it possible to use \nj{a} less expensive network.

\section{Conclusion}

In this paper, we proposed a new image enhancement method by a recurrently-trained super-resolution network. 
\nj{The proposed recurrent learning} consists of two phases: \nj{Network Training} phase to learn HR form LR, and \nj{Target Update} phase to apply \nj{the trained network} to the original to produce HR \nj{image for} the next stage. We clarified the \nj{characteristics} of recurrent learning by analyzing the process and results. Using image difference, we can show the results converge to a specific improvement point without divergence. Also, by using VIQET mean-opinion-score, it has been \nj{numerically shown that the quality of an image improves clearly as we repeat the learning procedure.} 
It is the first time to use these objective \nj{measures} in the image-enhancement \nj{area}, \nj{which is meaningful in} that the MOS score \nj{is useful} not only in a theory \cite{1248017} but also in practice. 

\nj{The DIV2K image dataset used in this paper has an equivalent image quality to that of an} actual digital broadcasting video, \nj{which is different from the conventional datasets with low quality used in the existing super-resolution papers.} Nonetheless, by RTS, the MOS score increased more \nj{than} the difference between the original and its low-resolution version, which is impossible with the conventional one-time learning. The increased MOS score proves that image enhancement through recurrent learning is efficient.

In order to clarify the \nj{effectiveness of the proposed RTS}, we experimented with various networks. The results show that recurrent learning can make an additional improvement, regardless of the  network used. And the results show a way to \nj{obtain a} cheap but effective network. \nj{We expect that it will contribute to the related society by dramatically reducing the cost of super-resolution and image enhancement.}

{\small
\bibliographystyle{ieee}
\bibliography{egbib}
}

\newcommand{\beginsupplement}{%
        \setcounter{table}{0}
        \renewcommand{\thetable}{S\arabic{table}}%
        \setcounter{figure}{0}
        \renewcommand{\thefigure}{S\arabic{figure}}%
     }
     
\beginsupplement

\end{document}